\documentstyle[twocolumn,aps]{revtex}

\input psfig.sty

\def\ltsim{\vbox {\hbox{\lower 0.6\baselineskip \hbox{$<$}} \break
		 \hbox{\lower 0.1\baselineskip \hbox{$\sim$}} } }
\def\gtsim{\vbox {\hbox{\lower 0.6\baselineskip \hbox{$>$}} \break
                 \hbox{\lower 0.1\baselineskip \hbox{$\sim$}} } }
\def\k{{\bf k}}

\def\r{{\bf r}}
\def\H{{\bf H}}
\def\j{{\bf j}}

\def\vs{{\bf v}_s}
\begin{document}
\draft

\twocolumn[\hsize\textwidth\columnwidth\hsize\csname %
@twocolumnfalse\endcsname

\title{Quasiparticle Transport in the Vortex State of Unconventional
Superconductors}

\author{
 P.J. Hirschfeld
}

\address{
Department of Physics, University of Florida, Gainesville, FL 32611, 
USA.\\
}

\maketitle
\begin{abstract}
I consider the problem of the vortex contribution to quasiparticle
transport in unconventional superconductors with line nodes, and argue that
the magnetic field dependence of transport coefficients is fixed
by the same scattering processes which limit low-temperature transport. 
I give no detailed calculations, but show that qualitatively
correct results may be obtained in the limit of low temperatures
and fields by simple physical arguments, based on estimates of the 
density of states and relaxation time in analogy with the zero-field
case.  I conclude with a brief discussion of the problem of anisotropy of
the field dependence  and influence of experimental  geometry. 

\end{abstract}

\pacs{PACS Numbers: 74.25.Fy, 74.72.-h,74.25.Jb}
]
{\it Introduction.} 
For some time, transport measurements in the Abrikosov vortex state of
cuprate superconductors have exhibited magnetic field and temperature
dependences which seemed difficult to incorporate into the standard
picture developed for classic superconductors.\cite{classictransport}  In
this model, if vortices are pinned, quasiparticles carry all electrical
current and entropy, and their transport is limited exclusively by
scattering from vortex cores.  The cores play a crucial role because of a
relatively high density of bound states created there by the local order
parameter supression, and because extended quasiparticle states are
completely gapped. 
\vskip .2cm
 In the past three or four years, a consensus has been established that the
order parameter in the hole-doped cuprate superconductors has $d$-wave
symmetry.\cite{Reviews}  It is therefore natural to ask if the new symmetry
class, and in particular the existence of gap nodes on the Fermi surface,
can invalidate the ``vortex scattering" model used to analyze quasiparticle
transport in classic superconductors.  In this paper, a summary of two
lectures given at the February 1998 Workshop on Strongly Correlated Systems
of the Asia Pacific Center for Theoretical Physics, I argue that this is
in  fact the case: that the existence of low-energy quasiparticle states
due to the order parameter nodes allows one to neglect vortex scattering at
small temperatures and fields, and focus instead on the rearrangement of
nodal quasiparticle occupation numbers occasioned by the Doppler shifts of
quasiparticle energies in the superflow field of the vortex lattice. 
Relevant scattering processes are the same which dominate zero-field
transport, leading to the possibility that the successful ``dirty $d$-wave"
theory of the $H=0$ superconducting state transport in HTSC can be
extended to the vortex state without the introduction of further
parameters.  Predictions of the theory for the field dependence of, e.g.,
the thermal conductivity in the vortex state, are consistent with
measurements made to date at higher temperatures. At low temperatures, where
the predictions are drastically different from those of the vortex scattering
model, preliminary experimental results support the theory, but more
experiments are needed.
\vskip .2cm
   I begin by providing the qualitative
background necessary to understand the temperature dependence of $H=0$
transport coefficients in the cuprates.  I then summarize the arguments
why extended quasiparticle states may be assumed to dominate the vortex
bound states, and discuss the semiclassical approximations used to treat
them.  I show that at $T=0$, transport coefficients {\it increase}
quasilinearly  with field, but at finite temperatures exhibit a
nonmonotonic field dependence.   For the most part  I  restrict myself in
this paper to  qualitative arguments which can be shown to provide  the same
results as obtained by a more complete treatment; in doing so I am mostly
summarizing work with C. K\"ubert which has  appeared
elsewhere.\cite{KH,KHSNS,KHtransport}  However, in section 4, I consider
the influence of  relative configuration of magnetic field  and current,
and give as a concrete example new  results for thermal transport as a
function of field in the basal plane of a
$d_{x^2-y^2}$ superconductor.  
\vskip .2cm
\section{ Introduction: zero-field transport}
At sufficiently low temperatures, the quasiparticle transport time will be
determined by impurity scattering.  If one restricts one's attention to the
$s$-wave scattering channel, but assumes an unconventional superconducting state
in which $\sum_k \Delta_k=0$, Pethick and Pines \cite{PethickPines} showed
in the context of heavy fermion superconductivity that $1/\tau(\omega )$ 
varies as $N(\omega)$ in the conventional weak scattering (Born) limit, and
as $N(\omega)^{-1}$ in the strong scattering (unitarity) limit.  The 
strength of individual scattering events therefore has profound consequences
for transport coefficients $L$, which may be estimated as
\begin{eqnarray}
L&\sim &\int \left( {-\partial f\over \partial \omega} \right) N(\omega )
\tau (\omega ) \sim N(T) \tau (T)\\
&\sim&\left\{ \begin{array}{ll} N(T)^2 & unitary \\
L_N\simeq \mbox{const.} & Born
\end{array}
\right.\nonumber
\end{eqnarray}
where $L_N=L(T_c)$.
\vskip .2cm
Experimentally, in both heavy fermion and cuprate superconductors, it is
found that in good samples, bulk transport coefficients are orders of magnitude
smaller than their normal state counterparts, thus leading to the (hitherto
microscopically unjustified) {\it ansatz} of unitarity scattering by simple
defects in these sytems.  Formally, such calculations are carried out either in
the kinetic equation approach of Pethick  and co-workers,\cite{Arfietal}
or in the self-consistent $t$-matrix approximation (SCTM).\cite{HVW,SMV}  In the
unitarity limit, both approaches yield similar results for temperatures and
energies above a scale $\gamma\sim \sqrt{\Delta_0\Gamma}$, where $\Delta_0$ is
the gap maximum and $\Gamma$ is the impurity 
scattering rate in the normal state.  At
low temperatures, only the SCTM yields true gapless behavior (finite residual
density of states $N(0)$), as well as the remarkable prediction of universal
transport,\cite{PALee}, recently confirmed by experiment in thermal
conductivity measurements on
$Zn$-doped YBCO.\cite{Tailleferuniversal} Briefly, this occurs because
zero-energy quasiparticle states acquire a residual broadening $1/\tau(0)\equiv
\gamma$ in the presence of impurities due to the nodes of the gap, and the
residual density of states is also found to scale with $\gamma/\Delta_0$.  Thus
$L$ is generically of order
$N_0(\gamma/\Delta_0)/\gamma\sim N_0/\Delta_0$, i.e. $L$ is independent of the
impurity concentration to leading order in the disorder $\Gamma$.  To give an
explicit example, in a $d_{x^2-y^2}$ state, $\Delta_k=\Delta_0\cos 2\phi$ over 
a model cylindrical Fermi surface, the thermal conductivity in zero applied 
magnetic field for heat current ${\bf j}_Q$ in the plane is approximately 
\begin{equation}
\frac{\kappa^{el}(T)}{\kappa_{00}}\simeq 1+{7\pi 
T^2\over 5\Delta_0\Gamma}
\left(\ln^2 {1.14\Delta_0\over T} +
{\pi^2\over 4}\right). 
\end{equation}
 where $\kappa_{00}/T\equiv \pi
N_0v_F^2/(6\Delta_0)$ is the universal thermal
conductivity.\cite{SunMaki,NH,Grafetal}
\vskip .2cm
At higher temperatures in the cuprates, the mean free path due to
inelastic scattering  becomes smaller than that due to impurity scattering.  DC
transport coefficients generically then show the following structure: a rise at
low temperatures from the universal value to a maximum at intermediate
temperatures, followed by a decrease to the normal state value at $T_c$,
typically an order of magnitude or so less than the peak in clean systems, but
an order of magnitude greater than the coefficients at the lowest temperatures.
(See Fig. 1) The existence of the maximum has been qualitatively understood for
some time as arising from the competition between a collapsing inelastic
scattering rate
$1/\tau_{inel}$
below
$T_c$ due to an opening of a gap in the spectrum of excitations responsible for
inelastic scattering, and the decrease of the number of quasiparticle carriers
at low temperatures.\cite{Romeroetal,Bonnetal}  At low temperatures there may
be problems with the details of this simple argument, due to the energy
dependence of the scattering rate due to impurities, $1/\tau_{imp}$ (as
discussed above), but the overall features are
well-described by models which add impurity and inelastic scattering rates
incoherently and include the collapse of the inelastic scattering below $T_c$
in some form.  My co-workers and I argued in \cite{HPS} that at
sufficiently low temperatures, when quasiparticle states are well-defined,
$1/\tau_{inel}$ should vary as $T^3$ for the simple reason that
electron-electron collisions in a normal Fermi liquid with constant particle
number leads to $T^2$, and there is one additional factor of the temperature
arising from the linear $d$-wave density of states.   This estimate is borne
out in a weak-coupling Hubbard model calculation of $1/\tau$,\cite{Quinlanetal}
which yielded excellent results in comparison to both low-frequency microwave
conductivity\cite{HPS} and thermal conductivity\cite{PH} measurements,
including disorder dependence, over the entire temperature range.
\begin{figure}[h]
\begin{picture}(150,220)
\leavevmode\centering\includegraphics{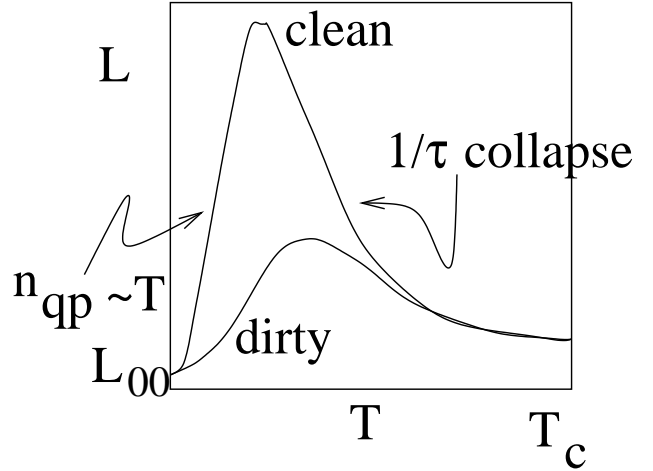}
\end{picture}
\caption{Schematic behavior of DC transport coefficients in the cuprates.
}
\end{figure}
\section{Quasiparticles in the unconventional vortex state}  I now turn to the
question of transport in the vortex state, focussing for consistency on
longitudinal thermal conductivity measurements, although microwave conductivity
measurements should 
share many of the same features.  Thermal conductivity as a probe
of the quasiparticle system has the 
advantage that in most cases of current interest
vortex motion may be ignored.  While the phonon contribution is often very
difficult to determine, it seems likely that it is field independent, so the
quasiparticle system is directly probed by measurement of the field dependent
part
$\delta\kappa(T;H)$.  This question is still open, however.  There are
several measurements of $\kappa$ at temperatures down to 10K or 20K in
YBCO-123 and BSCCO-2212, a typical recent one being that of
Freimuth,\cite{FreimuthPRL}  which shows at fixed temperture $T\sim 0.1 T_c$
that the thermal conductivity decreases with application of a field with
sublinear power or log behavior.  The qualitative behavior at higher
temperatures is not inconsistent with the conventional vortex scattering model,
in the sense that the quasiparticle mean free path shortens with increasing
field as might be expected if the vortices are moving closer together. 
However, the form of the field dependence is unusual, since, within the picture
just alluded  to, the number of scatterers should scale linearly with $H$. 
Furthermore, recent measurements at lower $T$ on both YBCO\cite{TailleferlowT}
and $UPt_3$,\cite{Suderowetal} a heavy fermion system also with an order parameter
with line nodes, have exhibited a thermal conductivity {\it increasing}
quasilinearly with field.  This can never be explained within a vortex
scattering picture.
\vskip .2cm
Vortex scattering in the cuprates is in fact likely to be negligible at low
fields for several reasons.  The  density of bound states in a state
with gap nodes was found to be significantly smaller than the density of
extended states by Volovik\cite{Volovik1} when averaged over the entire
vortex.  This calculation accounted for the gap nodes, but treated the bound
states as a continuum and thus must be regarded as an upper bound to the true
contribution.  In fact, in the cuprates the bound state separation or
``minigap" $\delta \sim \Delta_0 k_F^{-1}/\xi_0$ is very large compared to 
classic superconductors, and it is unlikely at low $T$ that more than a single
state/vortex is occupied.  The net cross section for quasiparticle scattering is
therefore likely to be very much smaller than in the classic case.  Finally, 
I note that at low temperatures the picture of independent scattering events by
vortices must break down in the perfect vortex lattice.  The
quasiparticle-vortex mean free path $\ell_{vortex}$ will then arise solely from
the  disorder-induced or 
residual thermally excited fluctuations of the
ideal periodic Abrikosov lattice.  At
$H=1T$, the intervortex distance of
$2R=\xi_0(2\pi)^{1/2}a^{-1}(H_{c2}/H)^{1/2}$, which 
provides an absolute lower bound for
$\ell_{vortex}$, is about 25 times the coherence length
$\xi_0$, or perhaps $400A$\, in 
HTSC.  While in YBCO the mean free path  due to defect
scattering and electron-electron collisions $\ell_{tr}$ at low T may
be as long as 3000A,\cite{Bonnetal}, in 
BSCCO $\ell_{tr}$  is a factor of ten or so
smaller,\cite{Morganetal} and therefore probably smaller than
$\ell_{vortex}$ over most of the field range of 
$10T$ or so in current experiments.  Note that this same argument 
applies to Andreev scattering by the vortex flow field as considered in
Ref. \cite{Yuetal}, since again vortex disorder, rather than the
intervortex distance, sets the mean free path.
\vskip .2cm
{\it If} mean free paths due to vortex scattering are indeed 
much longer than in classic
superconductors,   other scattering processes, such as
impurities and electron-electron collisions, will limit the mean free path at
low fields $H\ll H_{c2}$.   I will simply assume this for the discussion below,
and beg the question of the range of fields for which such an assumption may be
justified.
\vskip .2cm
In his seminal paper, Volovik showed that the dominant contribution of the
extended quasiparticle states to the density of states took the form
\begin{equation}
N(0;H)\sim N_0 \sqrt{H\over H_{c2}}.
\end{equation}
Such a term was indeed extracted from specific heat data on YBCO by Moler et
al.,\cite{Moleretal}  at the time a crucial piece of evidence in support of the
identification of $d$-wave symmetry in YBCO.  This result can be reproduced
easily if it is recalled that 1) the density of states in the pure $d$-wave
state varies as $N(\omega)\sim \omega/\Delta_0$ at low energies $\omega\ll
\Delta_0$; and 2) that a quasiparticle of momentum $\k$ moving in the superflow
field of a single vortex, $\vs({\bf r})\simeq \hbar/(2 m r)\hat \alpha$
($\alpha$ is the azimuthal angle in real space, e.g. in the plane if
$H\parallel\hat c$) experiences a Doppler shift $\omega\rightarrow \omega
-\vs\cdot \k$ in the lab frame.  We then have at $\omega=0$
\vskip .2cm
\begin{eqnarray}
N(0;H)\simeq \langle {|\vs\cdot \k|\over \Delta_0}\rangle_H \sim {1\over R^2}
\int_0^R dr\, r \left({\xi_0\over r}\right)
\end{eqnarray}
where in the second approximate equality an average $\langle...\rangle_H$ has
been performed over the vortex unit cell, of size $R$.  Since the integrand is
constant, $N(0;H)\sim R^{-1}\sim \sqrt{H}$ follows immediately.
\vskip .2cm
\section{Thermal conductivity at $H>0$}
Similar arguments can be applied\cite{KHtransport} to the kernel of the thermal
conductivity Kubo formula. At $T=0$, analogous to the discussion in Sec. 1, we
have 
\begin{eqnarray}
N(0;\vs)\sim v_s ~~~~~&;& \tau(\vs)\sim N(0;\vs) \sim v_s\\ 
\Rightarrow {\kappa\over T}&\sim& \langle v_s^2 \rangle_H \sim H \log H,
\nonumber
\end{eqnarray}
since the integrand now varies as $1/r$ rather than as a constant.  Note in
this case the average requires a lower cutoff, of order $\xi_0$, which reflects
our ignorance of how the extended quasiparticle wavefunctions vanish in
the core region.  Furthermore, we implicitly assumed in Eq. (5) that the local 
conductivity $\kappa(\r)$ depended only on $r$: this is true only for
the special case $\j_Q \parallel \H$.  A treatment of the opposite case,
$\j_Q\perp \H$ shows that the measured (averaged) $\kappa$ is much smaller
since once necessarily adds thermal resistances in series rather than in 
parallel in this case.\cite{KHtransport}   
A full numerical evaluation shows a behavior slightly more
linear in field than given by Eq. (5), due to higher contributions of order
$v_s^4$, which give rise to true linear terms when averaged.  
\vskip .2cm
At higher temperatures, terms decreasing in field arise due to 1) logarithmic
energy dependence of the unitarity limit scattering 
rate (neglected in Eq. (1)),
and 2) inelastic scattering.  The competition between increasing and decreasing
contributions leads in this case to a minimum in the field dependence, at
$H\sim T^2$ in the case of 1).  Determination of the exact form including
inelastic scattering has not yet been worked out in detail, but we know these
effects are required for the understanding of cuprates in zero field above about
1K or so.  Simple estimates show that the minimum in this more realistic case is
shifted out to fields higher than heretofore measured; one can thus via this
method understand both the magnitude and form of the (apparently) monotonically
decreasing field dependences reported in, e.g. Ref. \cite{FreimuthPRL}.
\begin{figure}[h]
\begin{picture}(150,200)
\leavevmode\centering\includegraphics{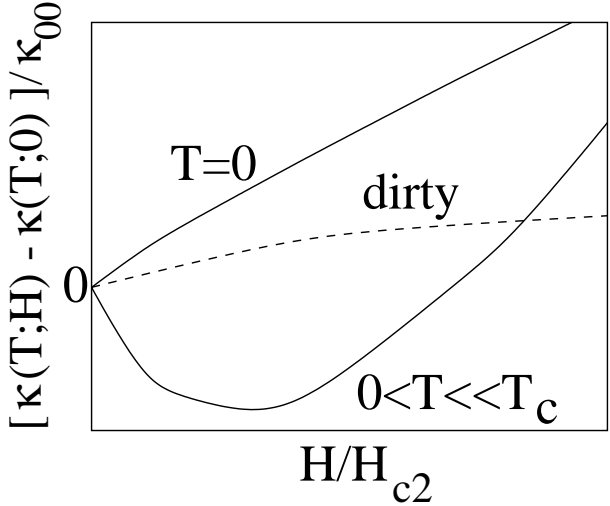}
\end{picture}
\caption{  Schematic behavior of $\kappa (T;H)/T$ 
at various temperatures for a
system with order parameter line nodes. Solid lines: clean system at $T=0$,
$0<T<T_c$; dashed line: dirty system at $T=0$.
 }
\end{figure}
\section{Anisotropy}
In the above estimates, I have supressed consideration of spatial angular
degrees of freedom in order to extract the qualitative field dependence.  In
general, the fourfold structure of the gap $\Delta_k$ in momentum space fixed
to the crystal field axes will induce an angular anisotropy in the current
response due to the coupling $\vs\cdot\k$.  On the other hand, I have already
mentioned in Sec. III a further source of anisotropy due to the different 
types of spatial averaging necessary for different directions of $\j_Q$ 
relative to $\H$. By itself, this leads generically to a maximum
in the thermal conductivity when $\j_Q\parallel H$, consistent with 
observations by Yu et al.\cite{Yuetal} and Aubin et al.\cite{Aubinetal}.
Unfortunately, this qualitative feature is also common to vortex scattering
theories.  It would be useful to consider an effect which is clearly
due entirely to the gap anisotropy.
\vskip .2cm 
  Such an example  is provided by taking
the thermal current $\j_Q$ parallel to
the field ${\bf H}$ in the plane, with  both making an angle $\epsilon$ with
respect to the $x$-axis.  In a coordinate system in which the quasiparticle
momentum is $\k=(\cos\phi,\sin\phi,0)$, the superfluid velocity is then
$\vs=(-\sin\epsilon\sin\alpha,\cos\epsilon\sin\alpha,\cos\alpha)$ (reminder:
$\alpha$ is the vortex winding angle). 
The quasiparticle energy shift is then
\begin{eqnarray}
\vs\cdot\k &=& v_sk_F \sin\alpha \sin(\phi-\epsilon)\\&\simeq& {v_sk_F\over
\sqrt{2}}\sin \alpha(\pm\cos\epsilon\pm\sin\epsilon),\nonumber
\end{eqnarray}
where the final approximate equality is justified becasue the kernel in the
Kubo formula is strongly peaked at the nodal momenta $\k_n$ with
$\phi=\pi/4,3\pi/4,...$.  Repeating the previous calculation, including the
\begin{figure}[h]
\begin{picture}(150,260)
\leavevmode\centering\includegraphics{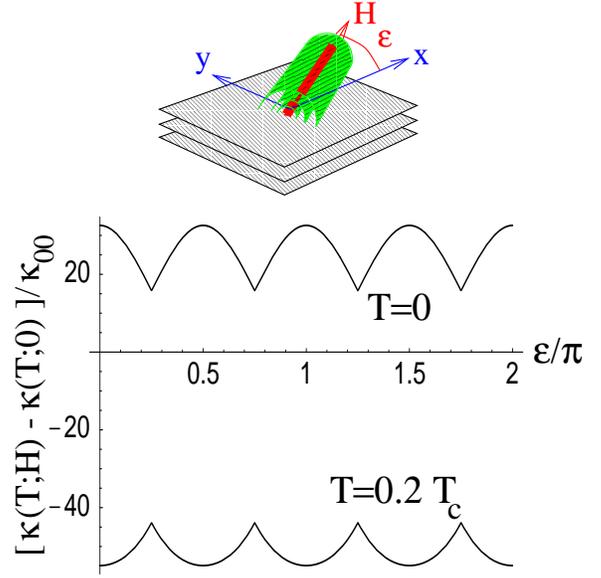}
\end{picture}
\caption{  Anisotropy of $[\kappa(T;H)-\kappa(T;0)]/\kappa_{00}$ as function of field direction $\epsilon$ (see insert) relative to the crystal axes for 
$T=0$ and $T=0.2T_c$ and $H/H_{c2}=0.01$.
Weak coupling value $\Delta_0/T_c=2.14$ was assumed
for the $d_{x^2-y^2}$ order parameter.
 }
\end{figure}
\noindent angular average, and remembering to symmetrize over all nodes, one finds (up to
factors of order unity)
\begin{eqnarray}
&&{\kappa(H)\over T}\sim \langle \left({\vs\cdot\k\over \Delta_0}\right)^2
\rangle_H \\
&&\sim {1\over R^2} \int_{\xi_{0}}^R dr\, r \left( {\xi_0\over
r}\right)^2
\int_0^{2\pi} d\alpha \sin^2\alpha
~\mbox{max}(\cos^2\epsilon,\sin^2\epsilon)\nonumber
\end{eqnarray}
so the thermal conductivity is seen at $T=0$ to 
exhibit 4-fold oscillations in the field
dependence  as a function  of field angle in the plane with a {\it mimimum} in the
nodal directions.    In Figure 3, I plot a full evaluation of the low-energy
approximation given in Ref.\cite{KHtransport} as a function of angle in the
basal plane.   Note that a traditional vortex scattering model would
predict no effect in this case.  At a higher temperature, the model of 
Ref. \cite{KHtransport} predicts a decreasing $\kappa(H)$ for the field
specified, and consequently {\it maxima} along the nodal directions; 
thus the oscillations appear
similar to those observed in an experiment
with slightly different geometry (but with $\bf H$ in the plane) by Yu et al.
\cite{Yuetal}, and attributed to Andreev scattering.    
A comparison of these two
effects is in progress.
\section{Conclusions}

I have tried to give a somewhat pedagogical summary of the ideas currently used
to interpret transport experiments at low temperatures in $d$-wave and other
unconventional superconducting states.   Back-of-the-envelope estimates for
the field dependence of thermal conductivity in a field were shown to
reproduce the more detailed calculations of recent work suggesting that the
field dependence of transport coefficients in heavy fermion and cuprate
superconductors can be well described by ignoring vortex scattering and
including only $H=0$ scattering processes, but shifting quasiparticle energies
in the vortex superflow field.  Fourfold oscillations of  transport
coefficients as a function of the angle between magnetic field direction
$\H\parallel\j$ and the crystal axes
 in the plane of 
a $d_{x^2-y^2}$ superconductor were
shown to be a natural consequence of this picture.
\vskip .2cm
 {\it Acknowledgments.}   Partial support was provided by
NSF-DMR-96--00105 and by the A. v. Humboldt Foundation.  The author is 
grateful to the Asia Pacific Center for Theoretical Physics for an
invitation to present these ideas in its 1998 Workshop on Strongly
Correlated Systems.

\end{document}